\documentclass[pre,preprint,groupedaddress,showpacs]{revtex4}
\usepackage{graphicx}
\usepackage{dcolumn}
\usepackage{amssymb,amsmath}
\DeclareMathOperator{\diag}{diag}
\DeclareMathOperator{\trid}{tridiag}
\DeclareMathOperator{\Ker}{Ker}
\DeclareMathOperator{\Imag}{Im}
\DeclareMathOperator{\rank}{rank}

\begin{document}

\title{Bifurcation Diagram for Compartmentalized Granular Gases}
\author{Devaraj van der Meer, Ko van der Weele, and Detlef Lohse}
\affiliation{Department of Applied Physics and J.M. Burgers Centre for
  Fluid Dynamics, University of Twente, P.O. Box 217, 7500 AE Enschede, The Netherlands}


\sloppy

\begin{abstract}
The bifurcation diagram for a vibro-fluidized granular gas in $N$ connected
compartments is constructed and discussed. At vigorous driving, the
uniform distribution (in which the gas is equi-partitioned over the
compartments) is stable. But when the driving intensity is decreased
this uniform distribution becomes unstable and gives way to a
clustered state. For the simplest case, $N=2$, this transition takes
place via a pitchfork bifurcation but for all $N>2$ the transition
involves saddle-node bifurcations. The associated hysteresis becomes
more and more pronounced for growing $N$. In the bifurcation diagram, apart
from the uniform and the one-peaked distributions, also a number
of multi-peaked solutions occur. These are transient states. Their
physical relevance is discussed in the context of a stability analysis.     
\end{abstract}

\maketitle

\section{Introduction}
One of the key features of a granular gas is the tendency to
spontaneously separate into dense and dilute regions
\cite{goldhirsch93,mcnamara94,du95,jaeger96,kudrolli97,kadanoff99}.
This clustering phenomenon manifests itself in a particularly clear manner
in a box that is divided in a series of $N$ connected compartments,
with a hole (at a certain height) in the wall between each two
adjacent compartments. The system is
vibro-fluidized by shaking the box vertically. With vigorous shaking the granular material is observed to be distributed uniformly
over the compartments as in any ordinary molecular gas. Below a
certain driving level however, the particles cluster in a small subset of the compartments, emptying all the others.

For $N=2$ the transition from the uniform to the clustered state is of
second order, taking place through a pitchfork bifurcation
\cite{eggers99}. For $N=3$ it was recently found that the transition
is hysteretic. It is a first order phase transition, involving
saddle-node bifurcations \cite{vdweele00}. This difference has been
explained by a flux model. In the present paper we will use the same
flux model to construct the bifurcation diagrams for arbitrary $N$. 

The main ingredient of this model is a flux function $F(n)$, which
gives the outflow from a compartment to one of its neighbors as a
function of the fraction of particles ($n$) contained in the
compartment \cite{eggers99}. The function $F(n)$ starts out from zero at $n=0$ and initially
increases with $n$. At large values of $n$ it decreases again because
the particles lose energy in the non-elastic collisions, which become
more and more frequent with increasing particle density. So $F(n)$ is
non-monotonic, and that is why the flux from a well-filled compartment can balance that from a nearly empty compartment.

\begin{figure}
\begin{center}
\includegraphics*[scale=1]{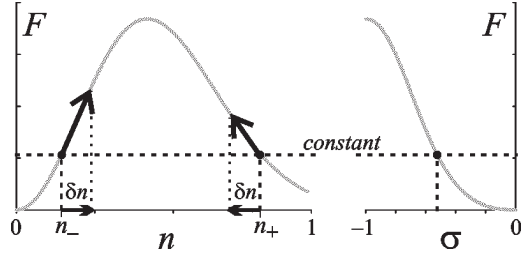}
\caption{The solutions $n_-$ and $n_+$ of $F(n_k)=constant$,
  cf. Eq.~\ref{eq-fixpoint}. Also  shown is how the flux balance
  responds to an increase of $n_-$ by an amount of $\delta n$ (see
  also Eq.~\ref{eq-netfl2}). The diagram on the right hand side
  depicts the relation between $F$ and the quantity
  $\sigma=F'(n_+)/F'(n_-)$, which plays an important role in the
  stability analysis of section 3. }
\label{fig-3}
\end{center}
\end{figure}

Assuming that the granular gas in each compartment is in thermal
equilibrium at any time (in the sense of the granular temperature \cite{mcnamara98})
the following approximate form for $F(n)$ can be derived \cite{eggers99}: 
\begin{equation}
\label{eq-grflux}
F(n_{k})=An_{k}^{2}e^{-BN^{2}n_{k}^{2}} ,
\end{equation}
which is a one-humped function, possessing the features
discussed before (See Fig.~\ref{fig-3}). In the above equation $n_{k}$ is the
fraction of particles in the $k$-th compartment, normalized to $\sum n_{k} =
1$. The factors $A$ and $B$ depend on the number of particles and their properties (such as
the radius, and the restitution coefficient of the
interparticle collisions), on the geometry of the system (such as the placement and form of the aperture between
the compartments), and on the driving parameters (frequency and
amplitude). The factor $A$ determines the absolute rate of the flux,
and will be incorporated in the time scale, which thus becomes
dimensionless. The clustering transition is governed only by $B$. 

The time rate of change $\dot{n}_{k}$ of the particle fraction in the $k$-th
compartment is given by the inflow from its two neighbours
minus the outflow from the compartment itself, 
\begin{equation}
\label{eq-netflux}
\dot{n}_{k}=F(n_{k-1})-2F(n_{k})+F(n_{k+1}) ,
\end{equation}
with $k=1,2,..,N$. Here we have assumed that the interaction is restricted to neighboring
compartments only. 

For a cyclic arrangement the above equation is
valid for all $N$ compartments (with $k=N+1$ equal to $k=1$). If we take non-cyclic boundary
conditions, by obstructing the flux between two of the compartments,
the equation has to be modified accordingly for these compartments.

The total number of particles in the system is conserved
($\sum_k n_k = 1$), so
\begin{equation}
\label{eq-totsum}
\sum_k \dot{n}_k = 0 .
\end{equation}

Statistical fluctuations in the system would add a noise term to
Eq.~(\ref{eq-netflux}), but we will not consider such a term here. So the
present analysis has to be interpreted as a mean field theory for the
system.
\\

Equation~(\ref{eq-netflux}) can also be
written in matrix-form, as
$\mathbf{\dot{n}}=\mathbf{M}\cdot\mathbf{F}$, or more explicitly:
\begin{equation}
\label{eq-ndot}
\begin{aligned}
\dot{n}_{k} &=\sum_{l}M_{kl}F(n_{l}) \\
&=
\left(
\begin{smallmatrix}
-2 & 1 & 0 & 0 & \cdots & 0 & 1 \\
1 & -2 & 1 & 0 & \cdots & 0 & 0 \\
0 & 1 & -2 & 1 & \cdots & 0 & 0 \\
: & : & : & : &  & : & :  \\
: & : & : & : &  & : & :  \\
1 & 0 & 0 & 0 & \cdots & 1 & -2
\end{smallmatrix}
\right) \cdot 
\left(
\begin{smallmatrix}
F(n_1)\\
F(n_2)\\
F(n_3)\\
:\\
:\\
F(n_N)
\end{smallmatrix}
\right)
\end{aligned}
\end{equation}

The given matrix $\mathbf{M}$ corresponds to a cyclic arrangement of
the $N$ compartments. A similar matrix can be written down for the
case of a non-cyclic arrangement. We will come back to this later, when
we will see that most of the results for the cyclic arrangement carry
over to the non-cyclic case.  

It is easily seen, from the fact that the elements of each row of $\mathbf{M}$
sum up to zero, that $\mathbf{1}=(1,1,\ldots,1)$ is an
eigenvector. The corresponding eigenvalue $\lambda=0$ physically reflects
the fact that the compartments cannot all be filled (or emptied)
simultaneously: $\sum_k \dot{n}_k = 0$ or $\sum_l M_{lk} = 0$. For
future reference we note that all the other
eigenvalues of $\mathbf{M}$ are negative (see Appendix).

The remainder of the paper is set up as follows. In Section~II we show
how to construct the bifurcation diagram, on the basis of
Eq.~(\ref{eq-ndot}), for an arbitrary number of compartments. In
Section~III we discuss the stability of the various branches in the
diagram. Section~IV discusses the physical consequences resulting from
the diagram, in particular in the limit for $N \to \infty$. Finally, Section~V
contains concluding remarks. The paper is accompanied by a
mathematical Appendix, in which some essential results concerning the stability analysis are derived.

\section{Constructing the bifurcation diagram}

To calculate the bifurcation diagram, we have to find the fixed
points of Eq.~(\ref{eq-ndot}) as a function of the parameter $B$,
i.e., those points for which $\dot{n}_k=\mathbf{M}\cdot \mathbf{F} = 0$. So
$\mathbf{F}$ must be a multiple of the zero-eigenvalue vector $\mathbf{1}=(1,1,\ldots,1)$. This tells us that, in a stationary
situation, all components
of the flux vector
$\mathbf{F}$ are equal: there is a detailed balance between all pairs of
neighboring compartments. This rules out, for instance, the possibility of
stable standing-wave-like patterns with equal but non-zero net fluxes
throughout the system. The fixed point condition now becomes
\begin{equation}
\label{eq-fixpoint}
\left\{ \quad
\begin{aligned}
F(n_k)&= \textit{constant} \\
\sum n_k &= 1
\end{aligned}
\right.      
\end{equation}

Since $F$ is a one-humped function, $F(n_k)=constant$ has two solutions, which will be called
$n_-$ and $n_+$ (see Fig.~\ref{fig-3}). Every fixed point
can be represented as a vector with elements
$n_-$ and $n_+$ (in any order, and summing up to 1) corresponding to a
row of nearly empty and well-filled compartments. Let us call the
number of well-filled compartments $m$. Apart from the
ordering of the elements, every fixed point is then specified by only two
numbers: $n_+$ and $m$. 

\begin{figure}[ht]
\begin{center}
\includegraphics[scale=1.2]{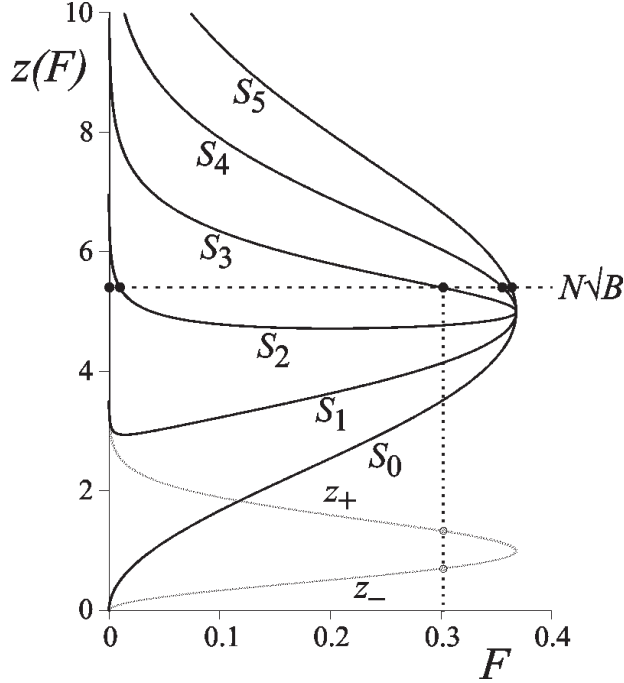}
\caption{Inverted flux functions $z_-(F)$ and $z_+(F)$ and the $N+1$
  sumfunctions $S_m(F)= m z_+(F) + (N-m) z_-(F), m=0,1,\ldots,N$. Here
  we picked $N=5$. The points of intersection with the horizontal
  line $z=N\sqrt{B}$ represent the fixed points for the parameter
  value $B$. Curves $S_0$ and $S_5$ correspond to the uniform distribution (below and
  above the critical point $B=1$, respectively) and the other curves
  belong to clustered states. Note that $S_0$ joins smoothly with
  $S_5$ at $B=1$ (i.e. $z=N\sqrt{B}=5$), and so does $S_1$ with $S_4$,
  and $S_2$ with $S_3$.}
\label{fig-1}
\end{center}
\end{figure}

Before actually calculating the bifurcation diagram, it is convenient
to replace the fraction $n$ by the (also dimensionless) variable $z=N n \sqrt{B}$, as then the
flux~(\ref{eq-grflux}) simplifies to $F(z_k)\propto z_k^2\exp(-z_k^2)$. The fixed point condition
Eq.~(\ref{eq-fixpoint}) then reads:
\begin{equation}
\label{eq-zfixpoint}
\left\{ \quad
\begin{aligned}
F(z_k)&= \textit{constant} \\
\sum z_k &= N \sqrt{B}
\end{aligned}
\right.      
\end{equation}      
So the $B$-dependence has been
transferred from $\mathbf{F}$ to the particle conservation, and this
enables us to
determine the entire bifurcation diagram from one single graph. This
is illustrated in Fig.~\ref{fig-1} for the case of $N$=5 compartments.

First, the one-humped function $F(z)$ is inverted
separately on both sides of the maximum, yielding the functions
$z_-(F)$ and $z_+(F)$. Then, we construct the sumfunctions:
\begin{equation}
\label{eq-sumf}
S_m(F)=m \, z_+(F) + (N-m) \, z_-(F)
\end{equation}

\begin{figure}[b]
\begin{center}
\includegraphics[scale=1.25]{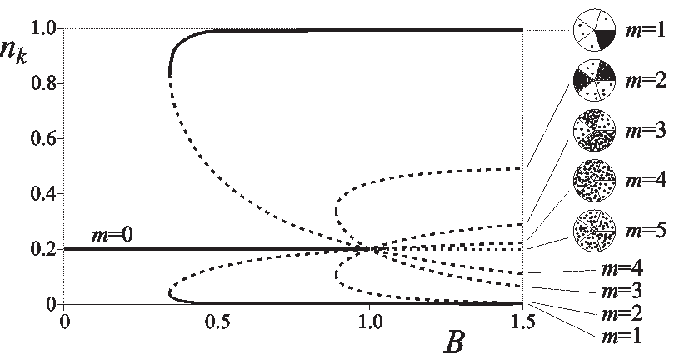}
\caption{Bifurcation diagram for $N=5$. It has been obtained from
  Fig. \ref{fig-1} by converting, for all $B$, each $\{z_-,z_+\}$-pair
  belonging to a point of intersection to a $\{n_-,n_+\}$-pair. Note
  that all branches come together at the critical point $B=1$. The
  (stable) $m=0$ branch becomes the (unstable) $m=5$ branch, the $m=1$
  branch turns into the $m=4$ branch, and $m=2$ switches to $m=3$.}
\label{fig-2}
\end{center}
\end{figure}

Now, from Eq.~(\ref{eq-zfixpoint}), the fixed points are found by
 intersecting the horizontal line
$z = N \sqrt{B}$ with the sumfunctions $S_m(F)$. In
Fig. \ref{fig-1} this is done for $B$=1.08. Each intersection point
 yields a pair $\{z_-,z_+\}$, or equivalently $\{n_-,n_+\}$. Repeating the procedure for all $B$, we obtain the
bifurcation diagram depicted in Fig. \ref{fig-2}.

It contains several branches. First, a
horizontal line (from the sumfunctions $S_0$ and $S_5$) corresponding
to the equal distribution $n_+=n_-=0.2=1/N$. Second, the
branches corresponding to the $m=1$ clustered state (from $S_1$),
which at $B=1$ goes over into the $m=4$ state (from $S_4$). And third,
the branch of the $m=2$ clustered state (from $S_2$), which at $B=1$
becomes the $m=3$ state. The physical appearance
of these solutions is sketched in the small diagrams. Note that only
the $m=0$ branch (i.e., the uniform solution up to $B=1$) and the
outer $m=1$ branch are stable. All
the other branches are unstable, as will be discussed in the next section.

At $B=1$, where the branches intersect with the uniform distribution
$n_+=n_-=1/N$, we have a critical point. In the flux function one
passes the maximum here. This means that $n_+$ and
$n_-$ are switched (relatively empty compartments become relatively
filled, and vice versa), so $m$-branches change into
$(N-m)$-branches. From a physical point
of view, the most important thing that happens at the $B=1$
intersection point is the destabilization of the uniform
distribution. 

The saddle-node bifurcations of the $m=1$ and $m=2$ branches
correspond to the minima of the sumfunctions $S_1$ and $S_2$
respectively, which in Fig~\ref{fig-1} can be seen to occur at $F
\approx 0.014$ for $S_1$ and $F
\approx 0.202$ for $S_2$. In general, if a sumfunction $S_m(F)$ has a
minimum for a certain $B$, the associated $m$ branch will have a
bifurcation. So the bifurcation condition is that the derivative
$dS_m(F)/dF$ equals zero, or equivalently:
\begin{equation}
\label{eq-bifpoint}
\frac{\left(\dfrac{dz_-}{dF}\right)}{\left(\dfrac{dz_+}{dF}\right)}=-\frac{m}{N-m}
\end{equation}      
Not surprisingly, the quantity on the left hand side
($dz_-/dz_+ \equiv \sigma$) will play an important role in the
stability analysis of the next section.

\section{Stability of the branches}
   
The stability of the branches (i.e., of the fixed points) is determined
by the eigenvalues of the Jacobi matrix $\mathbf{J}$ corresponding to Eq.~(\ref{eq-ndot}), with components:
\begin{equation}
\label{eq-jacobi}
J_{jk}= \frac{\partial \dot{n}_j}{\partial n_k} =
 \sum_{l}M_{jl}F'(n_{l}) \frac{\partial n_l}{\partial n_k} = M_{jk}F'(n_{k})
\end{equation}      
Here $F'$ denotes the derivative of $F$ with respect to $n$. Note that
the Jacobi matrix can also be written as the product of $\mathbf{M}$ and
the diagonal matrix $\mathbf{D}=\diag(F'(n_1),\ldots,F'(n_N))$, see
also Eq.~(\ref{eq-matr8}) in the Appendix. For a
fixed point the only diagonal elements that occur are $F'(n_+)$ ($m$
times) and $F'(n_-)$ ($N-m$ times), in any order. The ratio between
these two functions is precisely the quantity we encountered earlier in
the bifurcation condition Eq.~(\ref{eq-bifpoint}), namely $\sigma$:
\begin{equation}
\label{eq-sigma}
\sigma=\frac{F'(n_+)}{F'(n_-)}=\frac{dn_-}{dn_+}
\end{equation}
The Jacobi matrix $\mathbf{J}$ has $N$ eigenvalues, one of which is
always zero. The other $N-1$ eigenvalues depend on $m$ and the value
of $\sigma$.

For $m=0$ (the equipartitioned state) all non-trivial
eigenvalues are negative, up to the point $B=1$. This can be seen
either by direct numerical calculation, or analytically (see
Appendix). At $B=1$, the $m=0$ state becomes the $m=N$ state. Here,
the functions $F'$ in the Jacobi matrix~(\ref{eq-jacobi}) change sign,
and so do all of its eigenvalues. So suddenly the uniform state has $N-1$ \textit{positive} eigenvalues,
which implies a high degree of instability. Only in the limit $B \to
\infty$ does the uniform state regain some of the lost terrain: the
magnitude of all positive eigenvalues tends to zero here. Physically
speaking, in this limit the vibro-fluidization is too weak to drive
the particles out of the boxes anymore.  

\begin{figure}[ht!]
\begin{center}
\includegraphics[scale=1.41]{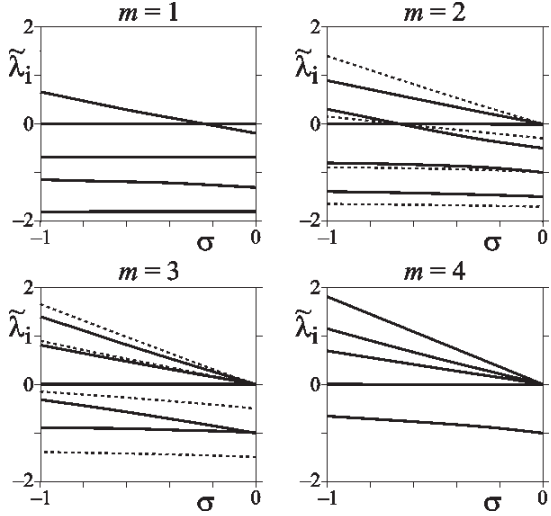}
\caption{Eigenvalues of the Jacobi matrix $\mathbf{J}$ as a function
  of $\sigma$, for the branches $m$ = 1, 2, 3, and 4. Rather than
  plotting $\lambda_i$, we display $\tilde{\lambda}_i =
\lambda_i/F'(n_-)$, because this yields a more clear-cut picture. Negative
  eigenvalues represent stable directions of the branches, and
  positive eigenvalues represent unstable ones. A zero
  crossing (such as for $m=1$ and $m=2$) indicates the occurrence of a
  saddle-node bifurcation. The value $\sigma=0$ corresponds to the
  limit $B\to\infty$, and $\sigma=-1$ to
  the critical point $B=1$. At this point, the eigenvalues of $m=1$
  and $m=4$ are equal but opposite in sign: the transition from the
  one branch to the other is marked by a distinct drop in
  stability. The same is true for the eigenvalues of $m=2$ and $m=3$,
  and also (not depicted) for those of $m=0$ and $m=5$. For $m$ = 2, 3
  there are two different cluster-configurations, with different
  eigenvalues. The dashed lines correspond to $\{++---\}$ for $m=2$,
  which goes over into $\{--+++\}$ for $m=3$. The bold lines apply to
  the slightly more stable configurations $\{+-+--\}$ and $\{-+-++\}$.}
\label{fig-eig}
\end{center}
\end{figure}

As for the
other values of $m$, in Fig.~\ref{fig-eig} we have plotted the
numerically evaluated eigenvalues (as functions of $\sigma$) for the system with
$N=5$ compartments. 

For $m=1$, we see that there are three eigenvalues that are always
negative.  The fourth non-trivial eigenvalue changes sign at
$\sigma=-0.25$. This corresponds to the saddle-node bifurcation of the
$m=1$ branch in the bifurcation diagram (Fig.~\ref{fig-2}), and the
bifurcation value of $\sigma$ is in agreement with Eq.~(\ref{eq-bifpoint}). The
region to the right of $\sigma=-0.25$ (where all non-trivial
eigenvalues are negative) belongs to the stable outer branch. The left part
$\sigma<-0.25$ belongs to the unstable inner branch, up to the point
$B=1$ (at $\sigma=-1$), where the $m=1$ branch goes over into the
$m=4$ branch. That is, the state $\{+----\}$ now switches to
$\{-++++\}$. At the same time all eigenvalues change sign, so
suddenly we have $3$ positive eigenvalues, which is only
one less than for the uniform $m=5$ state. (Indeed, the
only stable manifold  of the $m=4$ fixed point comes from the direction
of the completely unstable $m=5$ state). The positive eigenvalues never
cross zero anymore (there are no bifurcations beyond $B=1$) but, as
before, in the limit $B \to \infty$ ($\sigma \to 0$) they go to zero.  

For $m=2$ there are two possible configurations: $\{++---\}$ and
$\{+-+--\}$. Due to the cyclic symmetry, all other combinations are
equivalent to these two. The eigenvalues of the first configuration
are given by the dotted lines, and those of the second by the solid
lines. Although they are very similar (and are represented by exactly the same
branch in the bifurcation diagram), it is clear that the second
configuration is the more stable of the two. Apparently the two
well-filled compartments prefer to keep a distance.

The saddle-node bifurcation of the $m=2$ branch takes place at
$\sigma=-2/3$ [cf. Eq.~(\ref{eq-bifpoint})], where the third non-trivial
eigenvalue goes through zero. The fourth non-trivial eigenvalue always
remains positive, indicating that the $m=2$ branch never becomes
completely stable. (As a matter of fact, only the $m=0$ branch and part
of the $m=1$ branch can be completely stable). Note that for $\sigma
\to 0$ (large $B$) the positive eigenvalue tends to zero, so the
degree of instability is quite weak there. 

At $B=1$ the $m=2$ branch becomes the $m=3$ branch, with the two
configurations $\{--+++\}$ and $\{-+-++\}$, and with all eigenvalues
switching sign. As we see, the more dispersed configuration is again the
less unstable one. Also the phenomenon of all
positive eigenvalues going to zero as $\sigma$ approaches zero (the
weak driving limit $B \to \infty$) is again apparent.

\begin{figure}
\begin{center}
\includegraphics[scale=1.2]{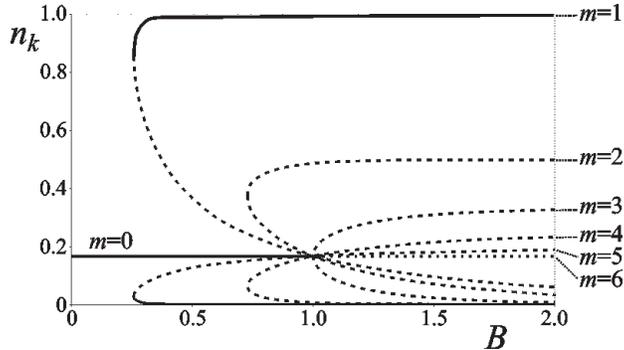}
\caption{Bifurcation diagram for $N=6$. Note the pitchfork bifurcation at $B=1$.}
\label{fig-bifN6}
\end{center}
\end{figure}

In the present example for $N=5$, and in fact for all odd values of
$N$, the branches in the bifurcation diagram are all born by means of
a saddle-node bifurcation. But for \textit{even} values of $N$ this is
different: in that case there is one branch-pair that springs from the uniform
distribution, at $B=1$, by a \textit{pitchfork} bifurcation. This is
illustrated in Fig.~\ref{fig-bifN6} for $N=6$. Here one sees all the
branches that were present already for $N=5$, only slightly shifted towards
the left, plus an additional pair of branches ($m=3$) bifurcating in the
forward direction from $B=1$.

The special status of the branch $m=N/2$ is also evident from
Eq.~(\ref{eq-bifpoint}), which tells us that the bifurcation condition
for this branch is $\sigma=-1$. This condition is fulfilled
only by $n_+=n_-=1/\sqrt{B}=1/N$. So, unlike all other
branches, this one originates at $B=1$ from the (until then stable)
uniform state. Related to this, the branch is the only one that is
symmetric for interchanging $n_+$ and $n_-$.

\section{Physical aspects}

The bifurcation analysis from the previous section can also be
understood from a more physical point of view. To this end, let us first have a closer
look at a 2-box system. In the equilibrium situation the net flux between the two boxes is
zero, with one filled ($n_+$) and one nearly empty ($n_-$) box. Suppose
the level of the empty box is raised by an amount $\delta n$. The
level of the filled box then decreases by an equal amount and the net
flux $\phi_{- \to +}$ from the empty to the filled box becomes
(see also Fig.~\ref{fig-3}):
\begin{equation}
\label{eq-netfl2}
\begin{aligned}
\phi_{- \to +} &= F(n_- + \delta n) - F(n_+ - \delta n) \\
&= \left( \frac{dF}{dn_-}+ \frac{dF}{dn_+} \right) \delta n = \left( 1 + \sigma \right) \frac{dF}{dn_-} \delta n   
\end{aligned}
\end{equation}
where we have used that $\sigma=dn_-/dn_+$ and
neglected the higher order terms in the Taylor expansion. There
are two different regimes. If $\sigma>-1$, 
the net flux is positive (as $F'(n_-)$ is always positive), so particles are flowing from $n_-$ to
$n_+$, restoring the equilibrium position. This is actually the
situation along the entire $m=1$ branch, for all $1<B<\infty$. For
$\sigma<-1$ (a situation which does not occur for our choice of $F$),
the net flux would be negative, raising the level of the emptier box even further,
 away from the equilibrium position. In the borderline case,
$\sigma=-1$ (at $B=1$), the system is indifferent to infinitesimal changes. 

This argument is readily generalized to the $N$-compartment system,
for an equilibrium with $m$ filled boxes. Now we raise the level of
all $N-m$ nearly empty boxes simultaneously by $\delta n$. This is
done by lowering all levels in the $m$ filled boxes by an equal
amount, which by particle conservation must be equal to $\delta n (N-m)/m$. The equivalent of Eq.~(\ref{eq-netfl2})
for the flux between any of the empty boxes to a neighbouring filled
box then reads:
\begin{equation}
\label{eq-netflux2}
\begin{aligned}
\phi_{- \to +} &= F(n_- + \delta n) - F(n_+ - \frac{N-m}{m}\delta n) \\
&= \left( \frac{dF}{dn_-}+ \frac{N-m}{m} \frac{dF}{dn_+} \right)
\delta n \\
&= \left( \frac{m}{N-m} + \sigma \right) \frac{dF}{dn_-} \delta n   
\end{aligned}
\end{equation}      
From this expression it follows that the transition between
a (relatively) stable ($\sigma>-m/(N-m)$) and a (relatively) unstable
($\sigma<-m/(N-m)$) configuration is marked by the bifurcation
condition Eq.~(\ref{eq-bifpoint}). So, by straightforward
physical reasoning we have reproduced the exact result obtained
earlier from an eigenvalue analysis.

The pitchfork bifurcation discussed at the end of Section~III is
especially important for $N=2$. In this case it is the \textit{only} non-uniform branch. To be
specific, it is a stable $m=1$ branch. This $N=2$ case \cite{eggers99}
is the only one
without any saddle-node bifurcations, and consequently it is the only
case where the change from the uniform to the clustered situation
takes place via a second order phase transition without any
hysteresis. For all $N>2$ the transition is of first order \cite{vdweele00}, and shows
a hysteretic effect that becomes more pronounced for growing $N$.

In the limit $N \to \infty$ the hysteresis is maximal: the first
saddle-node bifurcation takes place immediately after $B=0$, and this
means that there exists a stable $m=1$ solution over the entire range
$B>0$. So, if one starts out from this solution (at a certain
value of $B$) and then gradually turns down $B$, one will never
witness the transition to the uniform distribution. Vice versa, also the transition from the uniform solution to the
$m=1$ state will not occur in practice, even though the uniform
distribution becomes unstable at $B=1$. If one
starts out from the uniform solution (at a certain value of $B$ below
1) and increases $B$, one will witness the transition to a clustered
state, but in practice this will always be one with a number of
peaks. That is, the system gets stuck in a transient state with $m>1$,
even though such a state is not stable (it has one or more positive
eigenvalues).

The fact is that its lifetime may be exceedingly large, since the flux
in the neighborhood of a peak and its adjacent boxes (which are
practically empty) is very small. Furthermore, the communication
between the peaks is so poor that usually (even for moderate values of $N$) the dynamics comes to a standstill in a state
with peaks of unequal height.

\begin{figure}
\begin{center}
\includegraphics[scale=1.41]{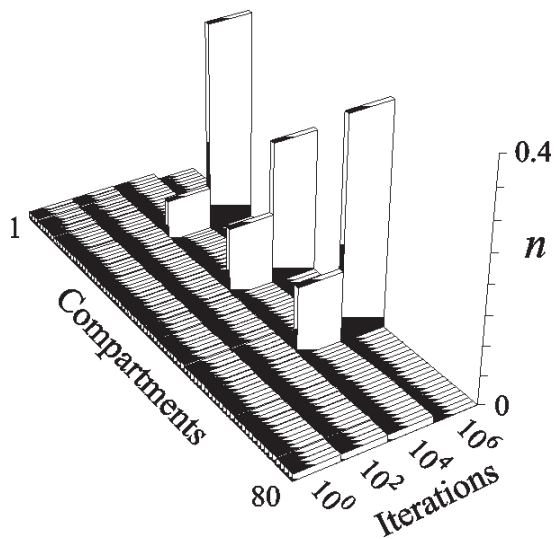}
\caption{Results from a numerical solution of Eq.~(\ref{eq-ndot}) for
  $N=80$, at $B=0.90$. Snapshots are taken after $10^0$, $10^2$, $10^4$ and $10^6$
  timesteps (iterations). Between 100 and 10,000 iterations a
  clustering pattern is seen take shape. Although strictly speaking
  this is a transient state, the system gets stuck in it.}
\label{fig-N80}
\end{center}
\end{figure}

Another point we would like to address is that practically the transition to a clustered state will take place already before $B=1$,
because the solution is kicked out of its basin of attraction by the
statistical fluctuations in the system \cite{vdweele00}. An example is shown in Fig.~\ref{fig-N80}. Here we see a
snapshot for the cyclic system with $N=80$ compartments, which were
originally filled almost uniformly, at $B=0.90$. The small random
fluctuations in the initial condition are sufficient to break away
from the (still stable) uniform distribution, and one witnesses the
formation of a number of isolated clusters. In the further evolution
these clusters deplete the neighbouring compartments and indeed the
whole intermediate regions. But the peaks themselves, once they are
well-developed, do not easily break down anymore. 

\section{Concluding remarks}

In this paper we have constructed the bifurcation diagram for a vibro-fluidized
granular gas in $N$ connected compartments. Let us now comment upon the result.

Starting out from $B=0$, i.e, vigorous shaking,
the equi-partitioned state is for some time the only (and stable)
fixed point of the system. For increasing $B$ we first come upon the
$m=1$ bifurcation, where the single-cluster state is born. For all
$N>2$ this happens by means of a saddle-node bifurcation, creating one
completely stable state and one unstable state (with $1$ positive
eigenvalue). The one with the largest difference between $n_+$ and
$n_-$ is the stabler one of the two states. Strictly speaking, there are $N$ equivalent
single-cluster states, since the cluster can be in any of the $N$ compartments.

For further growing $B$ we come across the $m=2$ bifurcation, where
two unstable 2-peaked states are created. The state with the largest
difference between $n_+$ and $n_-$ has $1$ positive eigenvalue, and the other one $2$. The two peaks can be distributed in
$\binom{N}{2}$ ways over the $N$ compartments, but as we have
seen they are not all equivalent. When the peaks are situated next to each
other we have a more unstable situation (the positive eigenvalues are
larger in magnitude) than when the peaks are further apart. This is
generally true for $m$-peaked solutions: of the
$\binom{N}{m}$ ways in which $m$ peaks can be distributed, the ones in
which the peaks are next to each other are the least favorable of all. 

\begin{figure}[htp!]
\begin{center}
\includegraphics[scale=1.41]{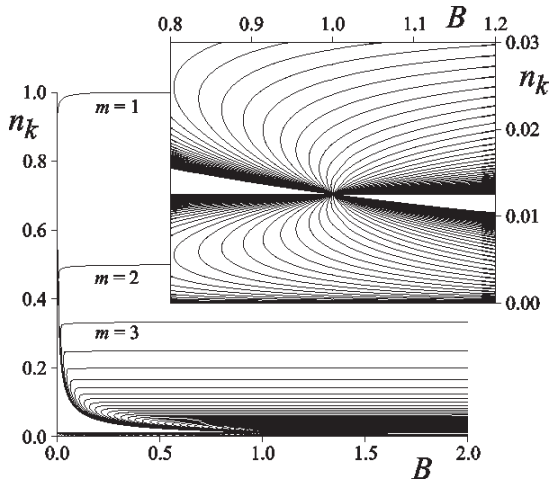}
\caption{Bifurcation diagram for $N=80$. The hysteresis extends almost
  all the way down to $B=0$, and there are numerous transient states
  (cf. Fig.~\ref{fig-N80}). The only strictly stable branches are the
  $m=0$ branch (up to $B=1$) at $n_k=1/N$, and the outer $m=1$
  branches. Naturally, the upper $m=1$ branch approaches $n_k=1$, the
  upper $m=2$ branch approaches $n_k=1/2$, the
  upper $m=3$ branch $n_k=1/3$, etc. The overlay picture shows the
  neighborhood of the critical point at $B=1, n_k=1/N$ in more detail.}
\label{fig-bifN80}
\end{center}
\end{figure}

For increasing $B$ we encounter more and more bifurcations, where
unstable $m$-clustered states come into existence (each
with $1$ more positive eigenvalue than the previous one), and for
large $N$ the bifurcation diagram is covered by a dense web of
branches. In Fig.~\ref{fig-bifN80} this is shown for N=80. The last saddle-node bifurcation takes place shortly before
$B=1$ and, for this even value of $N$, is followed by a final pitchfork
bifurcation (creating the $m=N/2$ branch) at $B=1$.  

The uniform solution (or $m=0$ state) is stable until $B=1$, with
$N-1$ negative eigenvalues and $1$ zero. For $B>1$ all its negative
eigenvalues become positive, making it suddenly the most unstable
state of all. Also, it now formally becomes the $m=N$ state. Moving
away from this uniform solution one encounters first the $m=N-1$
branch with $N-2$ positive eigenvalues, then the $m=N-2$ branch with
$N-3$ positive eigenvalues, etc. Finally, one arrives at the outermost
$m=1$ branch, which has no positive eigenvalues. This is the only
solution that is completely stable for $B>1$. But as we have seen in the
previous section, on its way from the uniform distribution to the
single peaked state, the system can easily get stuck in one of the
transient states (especially for large $N$) even though these are not
strictly stable.
\\

Throughout the paper, we have concentrated on the case where the $N$
compartments are arranged in a cyclic manner. But in doing so, we have in fact
also solved the non-cyclic case. Here we close the hole in the wall
between the $1$st and $N$th compartment, and consequently the flux
between them is zero. The matrix $\mathbf{M}$ then takes the following
form [differing from the cyclic one only in the first and last row, cf. Eq.~(\ref{eq-ndot})]:
\begin{equation}
\label{eq-matr5}
\mathbf{M}^{(nc)}=
\left(
\begin{smallmatrix}
-1 & 1 & 0 & 0 & \cdots & 0 & 0 \\
1 & -2 & 1 & 0 & \cdots & 0 & 0 \\
0 & 1 & -2 & 1 & \cdots & 0 & 0 \\
: & : & : & : &  & : & :  \\
: & : & : & : &  & : & :  \\
0 & 0 & 0 & 0 & \cdots & -2 & 1 \\
0 & 0 & 0 & 0 & \cdots & 1 & -1
\end{smallmatrix}
\right)
\end{equation}

The eigenvalue problem for this matrix is treated in the Appendix. One
eigenvalue is identically zero, and the other $N-1$ eigenvalues are
negative, just like for the cyclic system. This leads to a bifurcation
diagram that is indistinguishable from that of the cyclic case. Even
the stability along the branches is the same; only the magnitude (not
the sign) of the eigenvalues of the Jacobi-matrix $\mathbf{J}$ is slightly different for the two cases. 
\\

Finally, it should be emphasized that the results of the present paper
do not depend on the precise form of the flux function. We have
concentrated on the form given by Eq.~(\ref{eq-grflux}), but virtually
everything remains true for other choices of this function, as long as it is a non-negative, one-humped function,
starting out from zero at $n=0$ (no flux if there are no particles)
and going down to zero again for very many particles (no flux also in
this limit, since - due to the inelastic collisions -  the particles
form an inactive cluster, unable to reach the hole in the wall
anymore). Any function with these properties will produce a
bifurcation diagram similar to that of Eq.~(\ref{eq-grflux}).

In the likely case that the range of $\sigma=dn_-/dn_+$ is the
same, extending from $-1$ (this value is attained in the maximum) to
zero (in the outer regions of the flux function, for $n_-\sqrt{B}
\to 0$, $n_+\sqrt{B} \to \infty$), the bifurcation diagram will
have the same number of saddle-node bifurcations and the same number
of branches. The only things that change are the exact position of the bifurcation
points, and the magnitude of the eigenvalues along the branches.

Slight differences in the diagram would occur if the slope of $F$ on
the $n_+$ side was to become steeper than on the $n_-$ side. In that
case, the bifurcation condition Eq.~(\ref{eq-bifpoint}) would also have
solutions for $m>N/2$, thus allowing saddle-node bifurcations
for branches with $m>N/2$. These branches, however, would certainly be
quite unstable.

\appendix
\section*{APPENDIX: On the eigenvalues of $\mathbf{M}$ and $\mathbf{J}$}

In this appendix we present the analytical eigenvalues of the flux matrix
$\mathbf{M}$ [introduced in Eqs.~(\ref{eq-ndot}) and (\ref{eq-matr5})]
and discuss the eigenvalue problem for the Jacobian matrix
$\mathbf{J}$ [see Eq.~(\ref{eq-jacobi})], thereby determining
the stability of the branches in the bifurcation diagram.

First, we briefly treat the eigenvalues of $\mathbf{M}$. After that, we
turn to $\mathbf{J}$. In Subsection~2 we discuss its zero-eigenvalues: one eigenvalue is identically zero and, by
pinpointing the zero-crossing of a second eigenvalue, we reproduce the
bifurcation condition Eq.~(\ref{eq-bifpoint}). In Subsection~3 we determine the number of negative eigenvalues of $\mathbf{J}$ in the
low-driving limit $\sigma \to 0$. Likewise, in Subsection 4 we determine the
number of positive eigenvalues in the (mathematical) limit $\sigma \to
-\infty$. Combining these two results, in Subsection 5, we finally
find the number of positive eigenvalues of $\mathbf{J}$ for general
values of $\sigma$, and this gives the stability of the branches
over the entire bifurcation diagram.    

\subsection{Eigenvalues of matrix $\mathbf{M}$}    
The matrix $\mathbf{M}$ in Eq.~(\ref{eq-ndot}) is closely related to the $N \times N$
tridiagonal matrix $\trid(1,-2,1)$ associated with the second difference operator known from
numerical schemes for solving second order pde's. Its
eigenvalue problem can be solved exactly \cite{guardiola82}, and the
same is true for $\mathbf{M}$. The eigenvalues of $\mathbf{M}$ are given by:
\begin{equation}
\label{eq-eigval}
\lambda_{k} = -4 \sin^2 \left( \frac{k \pi}{N} \right)
\end{equation}
where $k$ runs from $0$ to $N/2$ for $N$ even, and from $0$ to
$(N-1)/2$ for $N$ odd. The corresponding eigenvectors are:
\begin{equation}
\label{eq-eigvec}
a_{i}(k) = C_1 \cos \left( \frac{(2i+1)k \pi}{N} \right) + C_2
\sin \left( \frac{(2i+1)k \pi}{N} \right) 
\end{equation}
with $i=1,\ldots,N$ and arbitrary coefficients $C_1$ and $C_2$.

As we see, the first eigenvalue ($k=0$) is zero and the corresponding
eigenvector is $\mathbf{1}=(1,1,\ldots,1)$. Physically, this eigenvector
represents simultaneous filling of all $N$ compartments, and the
eigenvalue $0$ expresses the fact that this is prohibited (because the
number of particles in our system is conserved).

All non-zero eigenvalues are negative and (except the one for $k=N/2$
in the case of even $N$) doubly degenerate. This means that the
corresponding eigenvectors span a two-dimensional subspace,
reflected by the two terms $C_1$ and $C_2$ in Eq.~(\ref{eq-eigvec}).
Since $\mathbf{M}$ is symmetric, and therefore normal, linear subspaces
corresponding to different eigenvalues are orthogonal. Especially, the
eigenvectors of all non-zero eigenvalues span a $N-1$ dimensional
subspace perpendicular to $\mathbf{1}=(1,1,\ldots,1)$. 

The matrix $\mathbf{M}^{(nc)}$ for the non-cyclic case, given by
Eq.~(\ref{eq-matr5}), has a different set of eigenvalues:
\begin{equation}
\label{eq-eigval2}
\lambda^{(nc)}_{k} = -4 \sin^2 \left( \frac{k \pi}{2N} \right)
\end{equation}
Here $k$ runs from $0$ to $N-1$. The corresponding eigenvectors are:
\begin{equation}
\label{eq-eigvec2}
a^{(nc)}_{i}(k) = \cos \left( \frac{(2i+1)k \pi}{2N} \right)
\end{equation}
Just like in the cyclic case, the first eigenvalue equals zero, and all
the others are negative. However, they are non-degenerate and the
corresponding eigenspaces are one-dimensional.

\subsection{Zero-eigenvalues of matrix $\mathbf{J}$} 
Now we turn to the Jacobian matrices. We consider the cyclic
version $\mathbf{J}$, with components as given in
Eq.~(\ref{eq-jacobi}), but the results are also valid for the
non-cyclic version. This matrix can be written as the product of
$\mathbf{M}$ and a diagonal matrix
$\mathbf{D}=\diag(F'(n_1),F'(n_2),\ldots,F'(n_N))$:
\begin{equation}
\label{eq-matr8}
\mathbf{J}=\mathbf{M}\cdot\mathbf{D}=
\left(
\begin{smallmatrix}
-2F'(n_1) & F'(n_2) & 0  & \cdots & 0 & F'(n_N) \\
F'(n_1) & -2F'(n_2) & F'(n_3) & \cdots & 0 & 0 \\
0 & F'(n_2) & -2F'(n_3) & \cdots & 0 & 0 \\
: & : & : &  & : & :  \\
: & : & : &  & : & :  \\
0 & 0 & 0  & \cdots & -2F'(n_{N-1}) & F'(n_N) \\
F'(n_1) & 0 & 0 & \cdots & F'(n_{N-1}) & -2F'(n_N)
\end{smallmatrix}
\right)
\end{equation}
 
In the context of the bifurcation diagram, the main thing one wants to
know is the number of positive eigenvalues of $\mathbf{J}$ for each
branch. This is what we are going to determine now.

First we note that the eigenvalues of $\mathbf{J}$ are real, even
though the matrix is not symmetric. This
is a consequence of the following similarity
relationship between $\mathbf{J}$ and $\mathbf{J}^{\dag}$:
\begin{equation}
\label{eq-app76}
\mathbf{J}^{\dag}=(\mathbf{M}\cdot\mathbf{D})^{\dag}=\mathbf{D}\cdot\mathbf{M}=\mathbf{D}\cdot(\mathbf{M}\cdot\mathbf{D})\cdot\mathbf{D}^{-1}=\mathbf{D}\cdot\mathbf{J}\cdot\mathbf{D}^{-1}
\end{equation}

This implies that $\mathbf{J}$ and $\mathbf{J}^{\dag}$ have
the same eigenvalues, and hence they must be real. Because
$\mathbf{M}$ is singular, $\mathbf{J}$ must be too (it has a zero
eigenvalue) and so its determinant $\det(\mathbf{J})$ is zero. More explicitly:
\begin{equation}
\label{eq-app1}
\det(\mathbf{J})= \det(\mathbf{M}) \cdot \det(\mathbf{D}) = \left( \prod_k F'(n_k) \right) \det(\mathbf{M}) = 0 
\end{equation}
where, for a fixed point with $m$ filled compartments, the product term equals $[F'(n_+)]^{m} [F'(n_-)]^{(N-m)}$.

For the other eigenvalues we have to look at the characteristic
equation $\det(\mathbf{J} - \lambda \mathbf{I})=0$. This is a
polynomial expression in $\lambda$, of which the constant term
is zero since it is equal to $\det(\mathbf{J})$. The
coefficient $L$ of the linear term is:
\begin{equation}
\label{eq-app2}
L = \sum_k \det(\mathbf{J}^{(k,k)}) = \sum_k \left( \prod_{l \neq k}
  \lambda_l \right)
\end{equation}
where the matrix $\mathbf{J}^{(k,k)}$ is the $(N-1)\times(N-1)$ matrix obtained from
$\mathbf{J}$ by deleting its $k$-th row and its $k$-th column. In the
right-hand side of this equation, the only product that survives is
the one that does
not contain the trivial (zero) eigenvalue. So:
\begin{equation}
\label{eq-app3}
L = \prod_{\text{all non-trivial} \lambda_l}
  \lambda_l 
\end{equation}
Alternatively, the determinant of $\mathbf{J}^{(k,k)}$ in Eq.~(\ref{eq-app2}) can be written in terms of
$\det(\mathbf{M}^{(k,k)})$, by deleting the $k$-th factor from the
product in Eq.~(\ref{eq-app1}): 
\begin{equation}
\label{eq-app4}
L= \sum_k \left( \prod_{l \neq k} F'(n_l) \right) \det(\mathbf{M}^{(k,k)})
\end{equation}

It can be shown that for all $k$ the determinant
$\det(\mathbf{M}^{(k,k)})$ is a constant,
$C$, which equals $(N-1)(-1)^{N-1}$ in the cyclic, and
$(-1)^{N-1}$ in the non-cyclic case. Thus, Eq.~(\ref{eq-app4}) reduces to:
\begin{equation}
\label{eq-app5}
L=  C \sum_k \left( \prod_{l \neq k} F'(n_l) \right) 
\end{equation}

For a fixed point with $m$ filled compartments, we can
write (using that in the above summation each of the products misses either an $F'(n_+)$ or an $F'(n_-)$):
\begin{equation}
\label{eq-app6}
\begin{aligned}
L &= C [F'(n_+)]^{(m-1)}[F'(n_-)]^{(N-m-1)} \times \\
& \quad \quad \quad \Big(  (N-m) F'(n_+) - m
  F'(n_-) \Big) \\
 &= C [F'(n_+)]^{(m-1)} [F'(n_-)]^{(N-m)} \Big( (N-m)\sigma - m  \Big)
\end{aligned}
\end{equation}
From this equation we conclude that $L$ becomes zero at $\sigma=
-m/(N-m)$. This is exactly the
bifurcation condition already given in the main text
[Eq.~(\ref{eq-bifpoint})]. Also, with Eq.~(\ref{eq-app3}), we see that an
eigenvalue crosses zero at this value of $\sigma$. 

It can
be shown, by a similar analysis, that the coefficient of the quadratic
term is not equal to zero at $\sigma=-m/(N-m)$, so not more than one
of the eigenvalues changes sign at the bifurcation.

\subsection{Number of negative eigenvalues of $\mathbf{J}$ for $\sigma \to 0$}
We now come to the next step in determining the number
of positive eigenvalues. We
again use the definition of $\sigma$ to write: $\mathbf{J}=F'(n_-)
\mathbf{M} \cdot \mathbf{\widetilde{D}}$, where
$\mathbf{\widetilde{D}}=\diag(1,\ldots,1,\sigma,\ldots,\sigma)$. The factors $1$ correspond to the $N-m$ nearly
empty boxes and the factors $\sigma$ to the $m$ filled boxes. The
precise ordering of the factors is not essential for the following
argument, so we may choose the above order for notational
convenience. 

The factor $F'(n_-)$ is always
positive, so we only have to deal with
$\mathbf{M}\cdot\mathbf{\widetilde{D}}$.
Note that only $\mathbf{\widetilde{D}}$ depends on $\sigma$ and that
in the limit $\sigma \to 0$ this matrix becomes\footnote{Since both the
  mapping $\sigma \to \mathbf{J}_{\sigma}$ as $\mathbf{J}_{\sigma}\to
  \det(\mathbf{J}_{\sigma}-\lambda \mathbf{I})$ are
  $\text{C}^{\infty}$ mappings (from $\mathbb{R} \to \mathbb{R}^{N
    \times N}$ and $\mathbb{R}^{N \times N} \to \text{P}(\lambda)$
  respectively, where $\text{P}(\lambda)$ is the space of
  polynomials of order $N$), it is allowed to take the limits
  $\sigma\to 0$ and $\sigma\to -\infty$, even though the latter does
  not occur in practice.}:
\begin{equation}
\label{eq-app7}
\lim_{\sigma \to 0} \mathbf{\widetilde{D}} = \diag(1,\ldots,1,0,\ldots,0) \equiv \mathbf{P}  
\end{equation}

$\mathbf{P}$ is a projection matrix which projects
$\mathbb{R}^N$ to the subspace spanned by the first $N-m$ unit
vectors. It is obviously non-singular, symmetric, and applying it twice gives the
same result as once: $\mathbf{P}^2=\mathbf{P}$. 

Instead of taking the
matrix $\mathbf{J}_0=\mathbf{M} \cdot \mathbf{P}$ as input for solving
our eigenvalue problem (in the limit $\sigma\to 0$), we will
rather look at the matrix $\mathbf{P}\cdot\mathbf{M}\cdot\mathbf{P}$
which is symmetric and has the same eigenvalues as $\mathbf{J}_0$. 

For proof of the last statement, let
$\mu$ be a (non-zero) eigenvalue of $\mathbf{J}_0$: $\mathbf{J}_0 \cdot
\mathbf{x}= \mu \mathbf{x}$. Then: $(\mathbf{P} \cdot \mathbf{M} \cdot \mathbf{P}) \cdot (\mathbf{P} \cdot
\mathbf{x}) = \mathbf{P} \cdot (\mathbf{M} \cdot \mathbf{P} \cdot
\mathbf{x}) = \mu (\mathbf{P} \cdot
\mathbf{x})$. Note that $\mathbf{P} \cdot \mathbf{x} \neq 0$, because otherwise also
$\mathbf{J}_0\cdot\mathbf{x}=
\mathbf{M}\cdot\mathbf{P}\cdot\mathbf{x}$ would be zero, contradicting
the assumption that $\mu$ is non-zero. This completes the proof.  

The matrix $\mathbf{M}$ is negative semi-definite. This means that
$\mathbf{M}$ has only negative or zero eigenvalues or, equivalently,
the inner product $\langle \mathbf{x},\mathbf{M} \cdot \mathbf{x} \rangle \leq 0$
for all $\mathbf{x}$. This means that also $\mathbf{P}\cdot\mathbf{M}\cdot\mathbf{P}$ is negative semi-definite,
because:
\begin{equation}
\label{eq-app9}
\langle \mathbf{x},\mathbf{P}\cdot\mathbf{M}\cdot\mathbf{P}\cdot
\mathbf{x} \rangle = 
\langle \mathbf{P}\cdot\mathbf{x},\mathbf{M}\cdot(\mathbf{P}\cdot\mathbf{x}) \rangle = 
\langle \mathbf{y},\mathbf{M}\cdot\mathbf{y} \rangle \leq 0
\end{equation}
In conclusion, $\mathbf{J}_0$ has negative and zero eigenvalues
only.

The remaining task is to identify the number of negative
eigenvalues, or otherwise stated, the rank of the matrix
$\mathbf{J}_0$. The statement which we shall prove is that
$\rank(\mathbf{J_0})=\rank(\mathbf{P})=N-m$.

Proof: Note that the image $\Imag(\mathbf{P})$ of $\mathbf{P}$ is
spanned by the first $m$ unit vectors of $\mathbb{R}^N$. Its kernel
$\Ker(\mathbf{P})$ is spanned by the remaining $N-m$ unit
vectors. Since the kernel of $\mathbf{M}$ is spanned by the vector
$\mathbf{1}$, the following identities hold:
\begin{subequations}
\begin{equation}
\label{eq-app10}
\Ker(\mathbf{P}) \cap \Ker(\mathbf{M}) = \mathbf{0} 
\end{equation}
\begin{equation}
\label{eq-app10b}
\Imag(\mathbf{P}) \cap \Ker(\mathbf{M}) = \mathbf{0}
\end{equation}
\end{subequations}

Now, for all $\mathbf{x} \in \Ker(\mathbf{P})$ it holds that $\mathbf{J}_0
\cdot \mathbf{x} = \mathbf{M}\cdot(\mathbf{P}\cdot\mathbf{x}) = \mathbf{0}$,
so $\Ker(\mathbf{P}) \subset \Ker(\mathbf{J}_0)$. On the other hand,
for all $\mathbf{y} \notin \Ker(\mathbf{P})$ one has
$\mathbf{P}\cdot\mathbf{y}\equiv\mathbf{z}\neq 0$, with
$\mathbf{z}\in\Imag(\mathbf{P})$, and therefore $\mathbf{J}_0
\cdot \mathbf{y} = \mathbf{M}\cdot \mathbf{z} \neq 0$ because of Eq.~(\ref{eq-app10b}). This means
that $\mathbf{y} \notin \Ker(\mathbf{J}_0)$, and thus
$\Ker(\mathbf{P}) \supset \Ker(\mathbf{J}_0)$. Together these two
results prove that $\Ker(\mathbf{P}) = \Ker(\mathbf{J}_0)$, so obviously the
rank of the two matrices must be equal. Since $\rank(\mathbf{P}) =
N-m$, this is also the rank of $\mathbf{J}_0$, which completes the proof.  

In short, we have shown that in the limit $\sigma \to 0$, the
Jacobi-matrix $\mathbf{J}$ has $N-m$ negative eigenvalues.

\subsection{Number of positive eigenvalues for $\sigma \to -\infty$}
We now turn to the limit $\sigma \to -\infty$. In this limit we
rewrite $\mathbf{J}$ as follows:
$\mathbf{J}=F'(n_-)\sigma\mathbf{M}\cdot\mathbf{\bar{D}}$. Here
$\mathbf{\bar{D}}=\diag(\sigma^{-1},\ldots,\sigma^{-1},1,\ldots,1)$,
which in the limit $\sigma \to -\infty$ becomes:    
\begin{equation}
\label{eq-app11}
\lim_{\sigma \to -\infty} \mathbf{\bar{D}} = \diag(0,\ldots,0,1,\ldots,1) \equiv \mathbf{Q}  
\end{equation}
Again, $\mathbf{Q}$ is a projection matrix, which now projects
$\mathbb{R}^N$ to the subspace spanned by the last $m$ unit
vectors, so $\mathbf{Q}$ is complementary to $\mathbf{P}$. Following
the same line of reasoning, but keeping in mind that now the
constant factor in front of $\mathbf{J}_{-\infty}$ is negative, we find that in
the limit $\sigma \to -\infty$, the
matrix $\mathbf{J}$ has $m$ \textit{positive}
eigenvalues.

\subsection{Number of positive eigenvalues of $\mathbf{J}$ for general
  $\sigma$}
We are now ready to draw the conclusion. Just below $\sigma=0$ the matrix $\mathbf{J}$ must, by continuity, have at least $N-m$ negative
eigenvalues. If we now move from $0$ towards $-\infty$, beyond a
certain point there must be at least $m$ positive eigenvalues (or
equivalently, at most $N-m-1$ negative
eigenvalues). We already know [cf. Eq.~(\ref{eq-app6})] that along the way exactly one
eigenvalue changes sign, at $\sigma=-m/(N-m)$. Taken together, this means that $\mathbf{J}$ has $m$ positive and $N-m-1$
negative eigenvalues for $\sigma<-m/(N-m)$, and $m-1$
positive and $N-m$ negative eigenvalues for $\sigma>-m/(N-m)$.

This completes the determination of the number of positive eigenvalues
for the various branches in the bifurcation diagram.    


\end{document}